\documentclass[11pt]{article}
\usepackage{amsmath,amsthm,amsfonts,amscd,eucal,latexsym,amssymb,mathrsfs, authblk}
\usepackage{amsxtra, calc, bbm}
\usepackage{epsfig}  
\usepackage[latin1]{inputenc}
\usepackage{verbatim} 

\usepackage{soul, cancel, color}
\def\cE{{\ca E}}

\def\cG{{\ca G}}

\def\cL{{\ca L}}

\def\cU{{\ca U}}

\def\bC{{\mathbb C}}           

\def\bR{{\mathbb R}}

\def\beq{\begin{eqnarray}}
\def\eeq{\end{eqnarray}}

\newcommand{\ca}[1]{{\cal #1}}         

\def\de{\delta}
\def\ep{\varepsilon}

\def\la{\lambda}

\def\om{\omega}
\def\ph{\varphi}

\newcommand{\Id}{\mathbbm{1}} 

\newcommand{\bk}{{\boldsymbol{k}}}

\newcommand{\bde}{\boldsymbol{e}}
\newcommand{\bm}{\boldsymbol{m}}
\newcommand{\bdM}{\boldsymbol{M}}
\newcommand{\bmu}{\boldsymbol{\mu}}

\numberwithin{equation}{section}

\def\vsp{\vspace{0.2cm}}
\def\vspp{\vspace{0.1cm}}

\def\sse #1 {\vsp\ifhmode{\par}\fi\refstepcounter{subsection}
  \noindent {\bf\thesubsection}. {\em #1}.\quad
  \addcontentsline{toc}{subsection}{\protect\numberline{\thesubsection} #1}%
  }

\def\ssb #1 {\vsp\ifhmode{\par}\fi\refstepcounter{subsection}
  \noindent {\bf\thesubsection.} {\bf #1.}\quad
  \addcontentsline{toc}{subsection}{\protect\numberline{\thesubsection} #1}%
  }

\def\ssa #1 {\ifhmode{\par}\fi\refstepcounter{subsection}
  \noindent {\bf\thesubsection.} {\bf #1.}\quad
  \addcontentsline{toc}{subsection}{\protect\numberline{\thesubsection} #1}%
  }

\def\remark #1 {\vsp\vspp\ifhmode{\par}\fi\noindent\noindent {\bf Remark.} {#1}\vsp\vspp\par}
\def\remarks #1 {\vsp\vspp\ifhmode{\par}\fi\noindent\noindent {\bf Remarks.} {#1}\vsp\vspp\par}

\title{Pale Glares of Dark Matter in Quantum Spacetime}
\author[1]{Sergio Doplicher}
\author[2]{Klaus Fredenhagen}
\author[3]{Gerardo Morsella}
\author[4]{Nicola Pinamonti}

\affil[1]{\small Dipartimento di Matematica, Universit\`a di Roma ``La Sapienza''\\

Piazzale Aldo Moro, 5, I-00185 Roma, Italy, email dopliche@mat.uniroma1.it}
\affil[2]{\small II Institute f\"ur Theoretische Physik, Universit\"at Hamburg\\

D--22761 Hamburg, Germany, email klaus.fredenhagen@desy.de}
\affil[3]{\small Dipartimento di Matematica, Universit\`a di Roma ``Tor Vergata''\\

Via della Ricerca Scientifica, I-00133 Roma, Italy, email morsella@mat.uniroma2.it}
\affil[4]{\small Dipartimento di Matematica, Universit\`a di Genova\\
Via Dodecaneso, 35, I-16146 Genova, Italy, 
and INFN Sezione di Genova, Italy, email pinamont@dima.unige.it}

\begin{document}

\maketitle

\begin{abstract}

A $\cU(1)$ gauge theory turns, on physically motivated models of Quantum Spacetime, into a $\cU(\infty)$ gauge theory, hence free classical electrodynamics is no longer free and neutral fields may have electromagnetic interactions. We discuss the last point for scalar fields, possibly describing dark matter;  we have in mind the gravitational collapse of binary systems or future applications to self gravitating Bose-Einstein condensates as possible sources of evidence of quantum gravitational phenomena. The effects so far considered, however, seem too faint to be detectable at present.
\end{abstract}

\section{Introduction}

One of the main difficulties of present day Physics is the lack of observation of quantum aspects of gravity; Quantum Gravity has to be searched without guide from nature, except the need to explain the observed universe as carrying traces of quantum gravitational phenomena in the only ``laboratory'' suitable to those effects,  the universe itself few instants after the Big Bang.

Looking forward to see those traces in the cosmic gravitational waves background (for the analysis of quantum linearized perturbations see the pioneering work \cite{MFB}, see also~\cite{BFHPR}), 
one can ask whether an expected consequence of
Quantum Gravity, the quantum nature of spacetime at the Planck scale, might leave observable traces.

Indeed Quantum Spacetime~\cite{DFR} would explain some aspects, as the horizon problem \cite{DMP}, usually explained by inflation, without having to make that hypothesis; but are there effects which $only$ QST would explain?

Free classical electromagnetism on Quantum Spacetime would be no longer free: the electromagnetic field and potential $F, A$ would fulfill
$$
\partial_\mu F^{\mu \nu}  -i [A_{\mu}, F^{\mu \nu}]  =  0,
$$
where the commutator would not vanish due to the quantum nature of spacetime.

This fact was noticed \cite{DF} at the very beginning of searches on Quantum Spacetime, as well as its first consequence: plane waves would be still solutions, but their superpositions would not in general, they would loose energy in favor of mysterious massive modes (see also~\cite{Z}). 

A naive computation showed that, by that mechanism, a monochromatic wave train passing through a partially reflecting mirror should loose, in favor of those ghost modes, a fraction of its energy - a very small fraction, unfortunately, of the order of one part in $10^{-130}$~\cite{DF}.  This looked too small to be worth a more accurate computation.

However QST should reveal itself, as discussed here below, causing an electromagnetic interaction of $neutral$ fields. This was noticed at the  beginning as well but looked even less promising of  visible consequences (see however~\cite{CJSWW}).

But the recent years brought increasing evidence of the role of dark matter, and the possibility of collapse of huge  dark matter binary systems; near the collapse, could those systems emit a seizable amount of electromagnetic radiation, and thus show a signature of the quantum nature of spacetime at the Planck scale?

In this note we discuss this problem, and show that a primitive, semiclassical  evaluation of that emission gives again a very small result: the fraction of the mass of such a system converted into electromagnetic radiation 
per unit time 
by the mechanism envisaged here would be less than  $10^{-89} s^{-1}$; nothing comparable to the few percents of the total mass converted into the gravitational wave radiation in the recently observed merges of binary black holes, GW150914 and GW151226, which inspire the numerical input of our calculation.

Our discussion proceeds as follows. In Sec.\ 1, after having recalled the main terminology, notations and results about the model of QST that we use, we discuss the action of the gauge group of QST on a neutral scalar field, and derive the interaction of the latter with the electromagnetic field by the covariant derivative prescription. Moreover, we show that such interaction can be described in terms of a magnetic moment associated to the scalar neutral field. Then, in Sec.\ 2, we evaluate the electromagnetic energy emitted in a state describing the precession of a stellar member of a collapsing binary system; that energy being computed, once the magnetic moment is evaluated, according to classical electrodynamics. We also comment on another manifestation of that magnetic moment, at first glance potentially giving rise to more visible effects, as they would be only quadratic in the Planck length (but hard to be detected anyway, see comments below): the electromagnetic (besides the gravitational) deviations of charged particles by a massive stellar object of dark matter interposed between us and a distant source. But in the case of the previously used data, we find a contribution to the angular deviation of  the order $10^{-34}$. \smallskip

\section{The magnetic moment of a neutral scalar field induced by Quantum Spacetime}
The model of QST adopted here is suggested by the Principle of $Gravitational$
$Stability$ $against$ $localization$ $of$ $events$ ~\cite{DFR}, ~\cite{BDMP}. This principle implies Spacetime Uncertainty Relations
\begin{equation}
\Delta q_0 \cdot \sum \limits_{j = 1}^3 \Delta q_j \gtrsim 1 ; \sum   
\limits_{1 \leq j < k \leq 3 } \Delta q_j  \Delta q_k \gtrsim 1 .
\end{equation}
for the coordinates $q_\mu$ of an event, which must be implemented by Spacetime commutation relations
\begin{equation}
       [q_\mu  ,q_\nu  ]   =   i \lambda_P^2   Q_{\mu \nu}
\end{equation}
where $\lambda_P$ is the Planck length and where $Q_{\mu \nu}$ satisfies appropriate Quantum Conditions.

The simplest solution is given by:
\begin{align}
[q^\mu , Q^{\nu \lambda}]  &=  0,\\
\label{dopleq14}
Q_{\mu \nu} Q^{\mu \nu}  &=  0,\\
((1/2) \left[q_0  ,\dots,q_3 \right] )^2 &= I,
\end{align}
where
\begin{align}
\left[q_0  ,\dots,q_3 \right]  &\equiv  \det \left(
\begin{array}{ccc}
q_0 & \cdots  & q_3 \nonumber\\
\vdots  & \ddots  & \vdots  \\
q_0 & \cdots  & q_3
\end{array}
\right)\\&\equiv  \varepsilon^{\mu \nu \lambda \rho} q_\mu q_\nu q_\lambda 
q_\rho =\nonumber\\&= - (1/2) Q_{\mu \nu}  (*Q)^{\mu \nu}
\end{align}
(notice that $Q_{\mu \nu} Q^{\mu \nu}$ is a scalar and $ Q_{\mu \nu}  (*Q)^{\mu \nu}$
is a pseudoscalar, hence we square it in the Quantum Conditions).

Called for brevity  \textit{the Basic Model of Quantum Spacetime}, this model implements 
exactly  the Space Time Uncertainty Relations and is fully Poincar\'e 
covariant. 

The $noncommutative$ C* algebra $\mathcal E$ of Quantum Spacetime can be associated to the above relations, by a procedure~\cite{DF, BDMP} which applies to more general cases. 

Assuming that the $q_\lambda,  Q_{\mu \nu}$ are selfadjoint operators and that the $Q_{\mu \nu}$ 
commute $strongly$ with one another and with the $q_\lambda$, the relations above can be seen as a bundle of Lie Algebra relations based on the joint spectrum of the $Q_{\mu \nu}$.

\textit{Regular representations} are described by representations of the C* group algebra of the unique simply connected Lie group associated to the corresponding Lie algebra, with the condition that $I$ is not an independent generator but is represented by the unit operator. They obey the Weyl relations
\begin{equation}\label{eq:weyl_rels} 
e^{ih_\mu q^\mu}e^{ik_\nu q^\nu}=e^{-\frac i2h_\mu Q^{\mu\nu} k_\nu}e^{i(h+k)_\mu q^\mu},\quad h,k\in\mathbb R^4.
\end{equation}

The C* algebra of Quantum Spacetime is the C* algebra of a continuous field of group C* algebras based on the spectrum of a commutative C* algebra.

In our case,  that spectrum - the joint spectrum of the $Q_{\mu \nu}$ - is the manifold $\Sigma$ of the real valued antisymmetric 2-tensors fulfilling the same relations as the $Q_{\mu \nu}$  do: a homogeneous space of the proper orthochronous Lorentz group, identified with the coset space of $SL(2,\bC)$ mod the subgroup of diagonal matrices.  Each of those tensors can be taken to its rest frame, where the electric and magnetic parts
 $\bde$,   $\bm$   are \textit{parallel unit vectors}, by a boost, and go back with the inverse boost, specified by \textit{a third vector, orthogonal to those unit vectors}; thus  $\Sigma$ can be viewed as the tangent bundle to two copies of the unit sphere in 3 space - its \textit{base}  $\Sigma_1$. 

Irreducible representations at a point of $\Sigma_1$ identify with \textit{Schr\"odinger's $p, q$ in $2$ degrees of  freedom}. The fibers are therefore the C* algebras of the Heisenberg relations in 2 degrees of freedom - the algebra of all compact operators on a fixed infinite dimensional separable Hilbert space.
 
The continuous field can be shown to be trivial. Thus the C* algebra  $\mathcal E$ of Quantum Spacetime is identified with the tensor product of
 the continuous functions vanishing at infinity on $\Sigma$ and the algebra of compact operators.

 The mathematical generalization of points are pure states. \textit{Optimally localized states} minimize
 $$\Sigma_\mu(\Delta_\omega q_\mu)^2;$$
  the minimum  being  $2$, reached by states concentrated on $\Sigma_1$, at each point  coinciding (if optimally localized at the origin)  with the   
  \textit{ground state of the harmonic oscillator}. Such states are the proper quantum version of points; the classical limit of Quantum Spacetime is then the product of Minkowski space and  $\Sigma_1$. Thus extradimensions, described by the doubled 2-sphere, are predicted by Quantum Spacetime. 
  
Optimally localized states are central in the definition of the  \textit{Quantum Wick Product}, which removes the UV divergences in the Gell-Mann--Low expansion of the S matrix for polynomial interactions on QST  ~\cite{BDFP:2003}.

The mentioned minimum, of the order of the squared Planck length in generic units, for the sum of the four squared uncertainties in the coordinates of an event is the first manifestation of a deeper fact: the minimum euclidean distance between two independent events in Quantum Spacetime is of the order of the Planck length in all reference frames. More generally, for each geometric operator like distance, area, three volume, or four volume, the sum of the squares of all spacetime components is, in each reference frame, at least of the order of the appropriate power of the Planck length~\cite{BDFP:2011}. 

These are mathematical results on the Quantum Geometry of Quantum Minkowski space. But dynamics, already at the level of a semiclassical treatment of Gravity, strongly suggests that the minimal distance between two independent events ought to have a dynamical meaning, $diverging$ when a singularity is approached~\cite{DMP}. This fact allows a possible solution of the horizon problem~\cite{DMP}, and will play a role in our discussion in the last Section.

Our first task is now to formulate and analyze gauge theories on the model of Quantum Spacetime just described. On ordinary classical spacetime, the gauge group of electromagnetism is the group of (regular) functions from Minkowski spacetime $\bR^4$ to $U(1)$, which can be regarded as (a subgroup of) the group of unitaries of the algebra $C_b(\bR^4) = M(C_0(\bR^4))$. Going to Quantum Spacetime, this should be naturally replaced by $\cG = \cU(M(\cE))$, the unitaries of the multipliers of the Quantum Spacetime algebra $\cE$. It is therefore a rather interesting possibility that the gauge group of electromagnetism could also act nontrivially on a real scalar field $\varphi(q)$ on QST, as
\begin{equation}\label{eq:gaugephi}
\varphi(q) \to U\varphi(q) U^*, \qquad U \in \cG.
\end{equation}
Of course on commutative spacetime the above action is instead trivial, because $U$ and $\varphi$ commute.

In order to find a Lagrangian invariant under the above action, we should introduce a covariant derivative $D_\mu$, i.e., a derivation on $\cE$ such that, under the action of $\cG$,
\[
D_\mu\varphi(q) \to UD_\mu\varphi(q)U^*.
\]
This is accomplished by defining
\begin{equation}\label{eq:covariantderivative}
D_\mu\varphi(q) := \partial_\mu \varphi(q) -ie[A_\mu(q),\varphi(q)],
\end{equation}
where $e$ is the electron charge (see below for a discussion of this choice) which describes the coupling with the gauge field, $\partial_\mu$ is the derivation on $\cE$ defined by
\[
\partial_\mu \varphi(q) = \frac{\partial}{\partial a_\mu} \varphi(q+a \Id)|_{a=0},
\]
and $A_\mu$ is the electromagnetic potential on QST, on which $\cG$ is assumed to act as
\begin{equation}\label{eq:gaugeA}
A_\mu(q) \to UA_\mu(q)U^* +ie^{-1}U\partial_\mu U^*,
\end{equation}
which reduces to the ordinary gauge transformation on commutative spacetime by writing $U = e^{i\Lambda}$. This also explains the choice of $e$ in~\eqref{eq:covariantderivative}, \eqref{eq:gaugeA} as the coupling constant between the electromagnetic potential and the neutral field $\varphi$. In fact, $A_\mu$ will also interact with the electron field $\psi$, which transforms as $\psi(q) \to U\psi(q)$, and therefore the choice $D_\mu\psi(q) = \partial_\mu\psi(q)-ieA_\mu(q)\psi(q)$ for its covariant derivative gives the correct interaction. A potential problem in this respect is represented by the fact that it seems difficult to write the interaction, on Quantum Spacetime, of $A_\mu$ with a field of charge different from $0,\pm e$ (like the quark fields). For a discussion in the framework of formal *-products and the Seiberg-Witten map see~	\cite{CJSWW}.

The fact that~\eqref{eq:covariantderivative} actually gives the correct definition of covariant derivative for the gauge transformation~\eqref{eq:gaugephi}, \eqref{eq:gaugeA} is easily verified: the transformed $D_\mu\varphi(q)$ reads in fact,
\[\begin{split}
\partial_\mu (U\varphi(q)U^*) &-ie[UA_\mu(g)U^*,U\varphi(q)U^*] + [U\partial_\mu U^*,U\varphi(q)U^*]\\
&=U D_\mu\varphi(q) U^* + \partial_\mu U \varphi(q) U^*  + U\varphi(q)\partial_\mu U^* + [U\partial_\mu U^*,U\varphi(q)U^*] \\
&=U D_\mu\varphi(q) U^*,
\end{split}\]
where the last equation follows from the fact that
\[
[U\partial_\mu U^*,U\varphi(q)U^*] = -\partial_\mu U \varphi(q) U^*-U\varphi(q)\partial_\mu U^*,
\]
which is easily verified using the commutativity of $U$ and $U^*$ and the identity $U\partial_\mu U^* = -(\partial_\mu U)U^*$, consequence of the Leibniz rule for $\partial_\mu$. 

We obtain therefore the following Lagrangian covariant under gauge transformations (which therefore gives rise to an invariant action)
\[
\cL = \frac12\eta^{\mu\nu}D_\mu \varphi(q)D_\nu \varphi(q) - \frac12m^2 \varphi(q)^2,
\]
and then, expanding the covariant derivatives, interaction terms between the neutral scalar field and the electromagnetic potential given by
\begin{equation}\label{eq:interaction}
\cL_I = -\frac{ie}{2}\{[A_\mu(q),\varphi(q)],\partial^\mu\varphi(q)\}-\frac{e^2}{2}[A_\mu(q),\varphi(q)][A^\mu(q),\varphi(q)],
\end{equation}
with curly brackets denoting the anticommutator. We note that on classical spacetime $\cL_I$ vanishes, as it should, since $A_\mu$ and $\varphi$ commute.\footnote{Note that $\cL_I$  would vanish also on Quantum Spacetime if the products appearing were interpreted as quantum Wick products~\cite{BDFP:2003}, as $E^{(n)}(f_1(q_1)\dots f_n(q_n))$ is independent from the ordering of the factors, and the tensor factors in $A_\mu(q)$ and $\ph(q)$ acting on Fock space would commute again.  But the Quantum Wick Product would anyway violate not only Lorentz invariance, but also Gauge Invariance, hence it could not be applied in the present context without first elaborating some radical modifications.}

Therefore, we will understand $\cL_I$ as defined through the noncommutative product in $\cE$. This has the drawback that it will depend explicitly on the center.

If we neglect, as customary, the quadratic term in $A_\mu$ (weak field approximation) we obtain, for the interaction part of the action,
\[
S_I = -\frac{ie}{2}\ \int d^4q\, \{[A_\mu(q),\varphi(q)],\partial^\mu\varphi(q)\} = -ie  \int d^4q\, A_\mu(q)[\varphi(q),\partial^\mu\varphi(q)],
\]
where the cyclicity of $\int d^4q$ was used. We obtain therefore the interaction of $A_\mu$ with a  current $j^\mu(q) = -ie[\varphi(q),\partial^\mu\varphi(q)]$, and the classical Euler-Lagrange equations for $A_\mu$ take the form
\[
\partial_\mu F^{\mu \nu} -ie[A_\mu, F^{\mu \nu}] = - j^\nu,
\]
with the field strength defined by $F_{\mu \nu} = \partial_\mu A_\nu - \partial_\nu A_\mu -ie[A_\mu,A_\nu]$ and transforming as $F_{\mu\nu} \to UF_{\mu\nu}U^*$.

In order to understand the physical meaning of this interaction, we now assume $A_0(q) = 0$, $A_h(q) = \frac12\ep_{hjk} B^jq^k$, the potential corresponding to an external constant magnetic field $\boldsymbol{B}$ in the classical spacetime limit ($\lambda_P \to 0$) and we again neglect in~\eqref{eq:interaction} the quadratic term in $A_\mu$, thus obtaining
\begin{equation}\label{eq:Binteraction}\begin{split}
\cL_I &= -\frac{ie}{4}\ep_{hjk}B^j\big\{\big[q^k,\ph(q)\big],\partial^h\ph(q)\big\} =\frac{e}{2}\ep_{jkh}B^jQ^{k\mu}\{\partial^\mu \ph(q), \partial^h  \ph(q)\},
\end{split}\end{equation}
where in the second equation we used the identity $[q^\nu, f(q)] = iQ^{\nu\mu} \partial_\mu f(q)$.

The above term corresponds then to the energy of a total magnetic moment $\boldsymbol{M}$ with components, in generic units, 
\begin{equation}\label{eq:magnetic}
M_j =  (e/2) \lambda _P ^2 \int_{q^0=t}d^3q\,\left[\frac12(\{\partial_l\ph,\partial^l\ph\}\de_{jk}-\{\partial_j\ph,\partial^k\ph\})m_k-\ep_{jkh}\{\partial_0\ph,\partial^h\ph\}e_k\right], \qquad j=1,2,3,
\end{equation}
where $e_k := Q^{0k}$ and $m_k := \frac 1 2 \ep_{khl} Q^{hl}$ are respectively the electric and magnetic components of the antisymmetric 2-tensor $Q^{\mu\nu}$. In the next section we will give some numerical estimates on the electromagnetic radiation and on the perturbations of the motion of charged particles associated to such a magnetic moment in suitable astrophysical situations.

\section{Some potentially observable consequences}

Defining, as usual, the free scalar field on QST as~\cite{DFR}
\[
\ph(q) = \int_{\bR^3} \frac{d\bk}{\sqrt{\om(\bk)}} [a(\bk)\otimes e^{-ikq}+a(\bk)^*\otimes e^{ikq}],
\]
and specializing to a point in the spectrum of the $Q$'s, where $\boldsymbol{e} =\boldsymbol{m}$, a computation yields the following expression for the total magnetic moment~\eqref{eq:magnetic}
\[\begin{split}
\bdM(t) = \frac{e\la_P^2}{2}\int_{\bR^3} \frac{d\bk}{\om(\bk)}&\left\{ \left[a(-\bk)a(\bk) e^{-2i\om(\bk)t}+a(\bk)^*a(-\bk)^* e^{2i\om(\bk)t}\right] \cos(\om(\bk)\bde\cdot\bk)\bk^2 \bde^\perp\right.\\
&\quad+\left.2a(\bk)^*a(\bk)[2\om(\bk)\bk\wedge \bde + \bk^2 \bde^\perp]\right\},
\end{split}\]
with $\bde^\perp = \bde - (\bk\cdot\bde)\bk/\bk^2$ the component of $\bde$ orthogonal to $\bk$. Therefore the effective magnetic moment of a particle with sharp momentum $\bk$ is given by
\begin{equation}\label{eq:momentsharp}
\bmu_{\bde,\bk} = e\la_P^2\left[2\bk\wedge \bde + \frac{\bk^2}{\om(\bk)} \bde^\perp\right].
\end{equation}

Of course detectable effects, if any, of the above interaction can be obtained in situations which give rise to a very large magnetic moment.  To this end, it is natural to consider a compact ``star'' of $\ph$ particles in rapid rotation around a very massive companion, akin to a binary pulsar or black hole. Neglecting the rotation of this star around its axis, a rough estimate of the associated magnetic moment $\bdM_S$ can be obtained by treating such an object as composed by classical particles in uniform rotation with a given angular frequency $\om$, and by associating to such particles the magnetic moment obtained from~\eqref{eq:momentsharp}. 

More in detail, we choose a reference system in which the orbit lies on the $(x,y)$ plane, and we indicate by $(\theta,\phi)$ the spherical coordinates of $\bde$ with respect to this system. Moreover, recalling that for the binary black hole giving rise to the event GW150914 the angular frequency just before the merger was $\omega \cong 471 \,s^{-1}$ and the radius of the orbit was $R \cong 350 \,km$~\cite{LV}, so that the speed in natural units ($\hbar = c = 1$) was $\beta = \om R \cong 0.6$, we may also assume that the motion of the particles is non-relativistic and approximate $\omega(\bk) \cong m$ in~\eqref{eq:momentsharp}. We obtain then
\begin{equation}\label{eq:starmoment}\begin{split}
\bdM_S(t) = e\la_P^2M&\left\{2R\om 
\left(\begin{matrix}
\cos \om t \cos \theta\\
\sin \om t \cos \theta\\
-\cos (\om t-\phi) \sin \theta
\end{matrix}\right)
\right.\\
&\left.+ R^2\om^2\left[
\left(\begin{matrix}
\sin\theta(\cos\phi-\sin \omega t \sin(\omega t-\phi)) \\
\sin \theta(\sin \phi + \cos \omega t \sin(\omega t-\phi))\\
\cos \theta
\end{matrix}\right)
+\frac 1 5 \left(\frac r R\right)^2
\left(\begin{matrix}
\sin\theta \cos\phi \\
\sin \theta \sin\phi \\
2\cos \theta
\end{matrix}\right)
\right]\right\},
\end{split}\end{equation}
with $M$ the mass of the object and $r$ its radius. In the particular case in which $\bde$ is normal to the orbital plane, $\bdM_S$ precedes then around it with the same angular frequency $\om$ of the object motion.

In the general case, $\bdM_S(t)$ can be written as sum of a constant moment, which of course does not give rise to emission of electromagnetic radiation, and of a time-dependent moment of the form
\[
\sum_{i} \bdM_i \cos(\omega_i t -\psi_i),
\]
with $\omega_i = \omega$, $2\omega$ and   $\psi_i$ are suitable phases. It is then an exercise in classical electromagnetism to verify that 
the time-averaged (classical) electromagnetic energy radiated (on classical spacetime) per unit time by this variable magnetic moment is given, in natural units, by
\begin{equation}\label{eq:power}\begin{split}
\frac{d \cE}{dt} &= \frac{2}{9}\sum_{\omega_i = \omega_j}\omega_i^4 ( \bdM_i \cdot \bdM_j) \cos(\psi_i-\psi_j)\\
&= \frac2 9 e^2\lambda_P^4 M^2 R^2\omega^6\left(1+\sin^2 \theta +\frac1 2 \omega^2 R^2 \sin^2 \theta\right).
\end{split}\end{equation}
Therefore averaging over the unknown direction of $\bde$ we get
\begin{equation}\label{eq:power2}
\frac{d \cE }{dt}   = \frac{8\pi}{27} e^2\lambda_P^4 M^2 R^2\omega^6 (5+ \omega^2 R^2) \simeq    e^2  \lambda_P^4 M^2 R^2 \omega^6  \simeq e^2  \left(\frac{\tau _P}{T} \right)^6 \left( \frac{R}{\lambda_P}\right)^2 M^2,
\end{equation}
where in the second equation we neglected numerical constants of order 1, and took into account that typically $\omega R \simeq 10^{-1}$ or smaller.
Taking then  $\cE \simeq M \simeq N m$, $N$ of the order of the number of particles in an object of the size of the sun and density of liquid water, i.e., roughly $10^{56}$,  $m \simeq 1 \,GeV$, the rotation period $T = 10^{-2} \,s$ and $R = 10^3\, km$ (comparable to the GW150914 parameters), and recalling that the Planck time $\tau_P \simeq 10^{-44}s$, we get that the fraction of energy radiated by the body per unit time is
\[
\frac{1}{\cE}\frac{d\cE}{dt} \simeq 10^{-89} s^{-1},
\]
and it is therefore negligible.

To make this sizable $T$ ought rather to be of Planckian order, which would probably mean that our object collapsed into a black hole and no radiation is visible, and anyway the above Minkowskian picture does not apply.

This computation is certainly too primitive,  but it suggests that the fraction of the total mass emitted as em radiation can be expected to be negligible;  by far nothing comparable with the fraction of a few percents emitted as GW in the BBH collapse GW150914.

But could a more cautious approach  {\it reverse} this conclusion? The question is legitimate, since  a heuristic argument, whose qualitative consequences are confirmed by a more cautious analysis~\cite{DMP}, suggests that near singularities  the {\it effective Planck length might diverge} as $\lambda_P  g_{00} ^{-1/2}$.

This might well introduce a metric dependent factor in our formula for the electromagnetic radiation caused by the magnetic moment of neutral matter, making it considerably larger in the last instants before the collapse into a black hole, heuristically as
$$
\frac{d \cE}{dt}      =   \frac{1}{g_{00}^{2}} e^2 \lambda_P^4 M^2 R^2 \omega^6,
$$
where $g_{00}$ is the time-time component of the background metric.

This qualitative conclusion is supported by the results in \cite{DMP}, which mean in particular that in a flat Friedmann- Robertson-Walker (FRW) background (which is spherically symmetric with respect to every point), with metric, in spatial spherical coordinates, $ds^2 = - dt^2 + a(t)^2[dr^2 + r^2dS^2]$, 
the size of a localisation region centered around an event at cosmological time $t$, measured by the radial coordinate $r$, must be at least of order $\lambda_P a(0) /a(t)$,  $t = 0$ being the time of the present epoch. 

The situation that we have in mind, namely that of a neutral object rotating in the gravitational field of a collapsing one, is of course better described by a Schwarschild metric  than by a FRW one. We note that the metric of a collapsing homogenous sphere of dust is given by the Oppenheimer-Snyder solution~\cite{OS,BS}, which is a Schwarzschild metric outside the sphere, matched with a closed FRW metric
\begin{equation}\label{eq:OSmetric}
ds^2 = -dt^2 + a(t)^2(d\chi^2 + \sin^2 \chi d\Omega^2)
\end{equation}
inside it. The scale factor $a(t)$ in the above metric can be expressed parametrically through the conformal time $\eta$

\begin{align}
a(t(\eta)) &= \frac12\sqrt{\frac{R^3_0}{2GM_0}}(1+\cos \eta), \label{eq:OSa}\\
t(\eta) &=  \frac12\sqrt{\frac{R_0^3}{2GM_0}}(\eta + \sin \eta), \label{eq:OSt}
\end{align}
with $M_0>0$ the ADM mass of the collapsing sphere and $R_0 \geq 2GM_0$ its initial areal radius. The conformal time at which the sphere is completely inside its Schwarzschild radius is given by $\eta_0 = \cos^{-1}(4GM_0/R_0 -1)$. 

The continuous match between the exterior Schwarzschild metric and the interior FRW one and the results of~\cite{DMP} recalled above seem therefore to justify the ansazt of replacing $\lambda _P$ in~\eqref{eq:power2} by $ \lambda _P a(0)/a(t)$. Indeed such an expression for the  effective Planck length converges to the usual value $\lambda_P$ in the limit $M \to 0$ in which the FRW metric becomes Minkowski, as one can easily verify by eliminating the conformal time $\eta$ from~\eqref{eq:OSa}, \eqref{eq:OSt}. Then, the above formula for the radiation of a precessing \emph{neutral} object would become
\begin{equation}\label{eq:effectivepower}
\frac{d \cE}{dt}      =  e^2  \left(\frac{\lambda_P a(0)}{a(t)} \right)^4 M^2 R^2 \omega^6.
\end{equation}
This energy has to be emitted of course at the cost of the kinetic energy of the rotating object due to spin, precession and orbital rotation, as well as of its potential energy, causing a faster inspiraling. For simplicity, we will  consider here only the orbital kinetic term, and then
\begin{equation}\label{eq:conservation}
\frac{d}{dt}\left(\frac12M R^2 \omega^2\right) = - \frac{d \cE}{dt},
\end{equation}
which entails that the total radiated energy can be estimated as
\begin{equation}\label{eq:totalenergy}
\cE = \frac12M R^2 \left[\omega_0^2 - \omega(t_{collapse})^2\right],
\end{equation}
where the integration cannot be extended beyond the hiding of our object within the event horizon of the other. For the radiated power according to our formulae near the singularity would diverge, but would  remain trapped and would not be visible from outside.

According to~\eqref{eq:totalenergy}, the total radiated energy can be sizable only if $\omega(t_{collapse}) \ll \omega_0$. In order to check if this is the case in typical situations, we solve~\eqref{eq:conservation}, that, inserting~\eqref{eq:effectivepower},
becomes
\[
\dot \omega = - e^2(\lambda_P a(0))^4M \frac{\omega^5}{a(t)^4},
\]
which can be integrated by separation of variables. To this end, we note that, by~\eqref{eq:OSt}, $dt/d\eta = a$, and therefore
\[
\int_{- \infty}^{t_{collapse}}  a(t)^{-4} dt = \int_0^{\eta_0} \frac{d\eta}{a(t(\eta))^3}= 8\left(\frac{2GM_0}{R^3_0}\right)^{3/2}\int_0^{\eta_0}\frac{d\eta}{(1+\cos \eta)^3}.
\]

Defining then, for $\eta \in [0,\pi)$,
\[
F(\eta) := \int_0^{\eta}\frac{dx}{(1+\cos x)^3} = \frac{\sin \eta(6\cos \eta+\cos(2\eta)+8)}{15(1+\cos \eta)^3},
\]
we obtain, neglecting numerical constants of order 1,
\begin{equation}\label{eq:omegacoll}
\omega(t_{collapse})^2 =  \left[\frac{1}{\omega_0^4}+e^2 \lambda_P^4M \left(\frac{R_0^3}{2GM_0}\right)^{1/2}F(\eta_0)\right]^{-1/2},
\end{equation}
which is smaller than $\omega_0^2$, as it should.

One can then observe that for $M_0 \to 0$ one has $\eta_0 = \cos^{-1}(4GM_0/R_0 -1) \sim \pi -\sqrt{8GM_0/R_0}$ and therefore $F(\eta_0) \sim \frac85(R_0/8GM_0)^{5/2}$, so that
\[
\omega(t_{collapse})^2 \sim \frac{G^{3/2}M_0^{3/2}}{e\lambda_P^2 MR_0^2  },
\]
would actually be very small with respect to $\omega_0^2$, making the total radiated energy~\eqref{eq:totalenergy} non negligible. 

 (Note that for ordinary matter the collapse would stop much before that the matter itself is hidden inside the horizon, due to the non vanishing pressure).

Moreover, this effect might disappear if one takes properly into account the red-shift of the radiation emitted near to the horizon. This could probably be done by using the general relativistic version of the radiated power by a magnetic dipole instead of~\eqref{eq:power}.

Conversely, for finite values of $M_0$, one can expand~\eqref{eq:omegacoll} due to the smallness of $\lambda_P^4$, and obtain for the total radiated energy
\[
\cE \simeq e^2\lambda_P^4M^2 R^2\omega_0^6 \left(\frac{R_0^3}{2GM_0}\right)^{1/2}F(\eta_0).
\]

Thus we see that for $2GM_0/R_0 =  1, F(\eta_0)$ vanishes, as it should, since the collapse takes place at the beginning. 
If instead $2GM_0/R_0$ is smaller than $1$ but of that order, then $F(\eta_0)$ is also of the same order: e.g., if  $2GM_0/R_0 = 1/2$ then $F(\eta_0) = 7/15$. Moreover, if we take as before $M_0 \simeq M \simeq 10^{56} GeV = 10^{37} M_P \simeq \cE _0$ and we recall that for $\hbar  = c = 1$ we have $G=M_P^{-2}$, we deduce $R_0  = 4GM_0 \simeq 10^{37} M_P^{-1} \simeq 10^{-1}\,km$, so that, assuming again $R \simeq 10^3 km \simeq 10^{41} M_P^{-1}$ and $T  \simeq 10^{-2} s  \simeq  10^{42} \tau_P$, we get
$$
\cE    \simeq e^2  \lambda_P^4 M^2 R^2R_0 \omega_0^6    \simeq  e^2  \lambda_P^4 M R^2R_0 \omega_0^6 \cE _0
 \simeq 10^{-96}  \cE _0 \simeq 10^{-40}\,GeV. 
$$

Note that using the same figures and multiplying the fraction of the total energy emitted as electromagnetic interaction per second, as given by the previous more brutal computation, Eq.~\eqref{eq:power2}, by the collapse time $t_{collapse} = t(\eta_0) = \frac1{\sqrt{2}}(\frac\pi 2-1)R_0 \simeq 10^{37} \tau_P \simeq 10^{-7} s$, we get an estimate of exactly the same order of magnitude.

Thus, as noticed earlier on in this discussion, the fraction of the mass converted into electromagnetic radiation is negligible, unless the period $T$ is at the Planck scale, which probably means that collapse took place, and that the emitted radiation is not visible to distant observers.  

As already mentioned, however,  a more realistic estimate ought to treat relativistically the electromagnetic emission. 

Eventually, another possibility, both more and less favorable, would be offered by a compact spinning concentrate of dark matter interposed to some distant source; spin and concentration apart, these objects exist  
and are revealed to us by gravitational lensing. Which results from the gravitational deflection of photons experimentally known since a century.

But if the source emits also charged particles, say electrons, sufficiently energetic to reach us within a reasonable delay after the $\gamma$ rays, their deflection ought to be modified by the magnetic field caused by the moment of our stellar object, due to Quantum Spacetime. A sort of QST - Northern Light phenomenon.

One might hope that this is a ``more favorable" situation with respect to the one considered above because, while the energy emitted is proportional to the fourth power of the Planck length, the deviation we are mentioning now would be only $quadratic$ in 
$\lambda_P$.

Nevertheless, a rough estimate of the deviation angle $\theta$ of an electron by a compact object of mass $M$ and radius $R$ spinning at angular velocity $\omega$ gives, using~\eqref{eq:starmoment},
\[
\theta \cong \frac{M_S}{m\gamma R^2} \cong \frac{e \lambda_P^2 M \omega}{m \gamma R},
\] 
with $m$ the electron mass and $\gamma = (1-v^2)^{-1/2}$. Choosing, as above, $M \simeq 10^{56} \,GeV$, $R \simeq 10^{-1}\, km$ and $\omega = 10^2 s^{-1}$ the deviation would be only $\theta \cong 10^{-34}$ for electrons of energy $1\, TeV$, which would reach us with a delay, with respect to photons, of a few hours if the source is $10^9$ light years distant. The delay for protons of $10^3 \,TeV$ (still considered to be lower than the GKW limit) would be the same, but the deviation would be $10^3$ times smaller. Of course the deviation would be more important for softer electrons, which however would reach the Earth when nobody is there any longer.

Moreover, a ``less favorable'' aspect  is that
electrons are considerably influenced by the much stronger galactic magnetic field, of which a precise knowledge would be needed, together with a nearly exact location in the sky of the sources of electromagnetic radiation and of electrons, as well as a clear recognition of the coincidence of their origin.

\section{Concluding remarks}
Our discussion was based on the choice~\eqref{eq:covariantderivative} of the covariant derivative, with $e$ denoting the electron charge. This choice seems to be dictated by gauge invariance in a theory which includes the electromagnetic interactions of the electron, taking into account the noncommutativity of $\cG = \cU(M(\cE))$, the group of unitaries in the multipliers of the algebra of Quantum Spacetime $\cE$. On an  $\cE$ bimodule, only the left (resp.\ right) action of $U$ (resp.\ $U^* $), or the trivial action, are allowed. 

This poses a problem for the Standard Model, apparently excluding quark fields. This problem has been noticed and discussed by several Authors,  see e.g.~\cite{CJSWW}. It deserves further discussions to see whether in our context the choice made in~\eqref{eq:covariantderivative} is really the only choice.

According to our preceding discussion,  so far there seems to be no indication of visible effects of the quantum nature of spacetime at the Planck scale, except its role in solving the horizon problem~\cite{DMP} and justifying from first principles part of the assumptions made in the inflationary scenario.

The effects considered in this note are so tiny that it would be instructive to compare them with those due to the graviton mediated dark matter - photon interaction.

Furthermore, the electromagnetic radiation emitted by a collapsing binary system due to the mechanism proposed here ought to be compared with the Hawking radiation.

But the QST induced electromagnetic interactions of dark matter might be detectable in more exotic hypothetic astrophysical objects, like self gravitating Bose-Einstein condensates of dark matter, consisting of neutral scalar particles. The stability of such objects, with a solar mass and a radius of few dozen of kilometers, has been recently investigated, both in the isotropic and rotating cases; the possible formation of vortices has also been considered (cf., e.g.,~\cite{MT}).

Smaller object of this nature were excluded in the quoted study by the nonrelativistic approximation used there, but might well be relevant to manifest sizable QST-electromagnetic effects, possibly also in the form of electromagnetic vortex-vortex interactions, which might potentially change the dynamics of these hypothetical objects. 

These points will be dealt with in subsequent studies.

\end{document}